\documentclass[12pt]{article}
\usepackage{amssymb}
\begin{document}
\begin{center}
{\bf Characteristic features of anharmonic effects in the lattice dynamics
of fcc metals} \vskip 5mm

M. I. Katsnelson$^1$, A. V. Trefilov$^2$, and K. Yu. Khromov$^2$.
\vskip 5mm

$^1$Institute of Metal Physics, Urals Branch of the Russian Academy of Sciences,
620219 Ekaterinburg, Russia

$^2$ Russian Research Center "Kurchatov Institute", 123182 Moscow, Russia
\vskip 10mm

{\bf Abstract} \vskip 5mm
\parbox{123mm}{
The dispersion in the entire Brillouin zone and the temperature dependence
(right up to the melting temperature) of the anharmonic frequency shift and
phonon damping in a number of fcc metals is investigated on the basis of
microscopic calculations. It is found that the anharmonic effects depend
sharply on the wave vector in the directions $\Gamma$-X, X-W, and $\Gamma$-L
and, in contrast to bcc metals, the magnitude of the effects is not due to the
softness of the initial phonon spectrum. It is shown that the relative
frequency shifts and the phonon damping near melting do not exceed 10-20\%.
The relative role of various anharmonic processes is examined, and the relation
between the results obtained and existing experimental data is discussed.
}
\end{center}
\vskip 5mm

The investigation of anharmonic effects (AEs) in lattice dynamics is a classic
problem of solid-state physics. It is important, specifically, because of the
role that such effects they can play in phenomena associated with structural
phase transitions and melting in crystals (see, for example, Refs. 1-3).
At the same time, obtaining any information about the magnitude and scale of
AEs from experiment and theory is a difficult problem. The experimental study
of such "basic" AEs as the frequency shift and damping of phonons is very
difficult and leads to a large uncertainty in the results (see, for example,
the data presented in Refs. 4 and 5 for bcc and fcc metals, respectively).
Up to now first-principles microscopic calculations of AEs have been performed
for one point of the Brillouin zone (N) in the bcc phase of Zr and four points
(N, P, $\omega$, G) in Mo. (Ref. 6)  Detailed information about AEs in the
entire
Brillouin zone and their temperature dependence has been obtained in Refs. 4
and 7 on the basis of pseudopotential theory for the bcc phases of alkali and
alkaline-earth metals. For these metals the most striking manifestations of
AEs are due to the "soft-mode behavior" (the anomalous temperature dependence
of the phonon frequencies) of the $\Sigma_4$ branch. It is of interest to
calculate AEs for the "general position", i.e., for crystals which do not
possess soft vibrational modes. Such crystals include most metals with
close-packed structures, for example, fcc. There is virtually no information
about AEs in such metals, not counting Ref. 5, where the damping of phonons in
precious metals and Al was calculated. However, the calculations were performed
only for individual points of the Brillouin zone and at room temperature (which
is much less than the melting temperature $T_m$). Moreover, the accuracy of the
model used in Ref. 5 for interatomic interactions is unclear, which is probably
the reason for the qualitative discrepancy between the theoretical and
experimental dependences of the damping on the wave vector. In the present
letter the AEs in the lattice dynamics of fcc metals are investigated in detail
on the basis of a systematic anharmonic perturbation theory.

The calculations were performed for Ir, the fcc phases of Ca and Sr, and
the hypothetical fcc phase of K. This choice of objects was determined by
the fact that reliable and at the same simple microscopic models which make it
possible to describe a wide range of lattice properties of these metals exist
Refs. 4,7-9. Iridium is an example of an fcc crystal with a "stiff interatomic
interaction potential, similar to the Lennard-Jones potential, Ref. 8
while K, Ca,
and Sr are characterized by potentials with a softer "core". To demonstrate the
characteristic features of AEs due to the specific nature of the potential
with the same lattice geometry, we present here the results for the "limiting"
cases --- Ir and K. The results for Ca and Sr are qualitatively similar to the
results for K and will be presented elsewhere.

The parameters determining the interatomic interaction potential are presented
in Ref. 8 for Ir and in Ref. 4 for K. The calculations were performed on the
basis of the anharmonic perturbation theory taking account of thermal expansion
(quasiharmonic contributions, qh) and three- and four-phonon interaction
processes to accuracy $\varkappa^2$, where $\varkappa= (m/M)^{1/4}$ is the
adiabatic smallness parameter, and m and M are the electron and ion masses.
The exact formulas actually used in the calculations are presented in Ref. 4.
For a qualitative discussion of the results, we present here their
high-temperature asymptotic representation for $T>\Theta_D$ ($\Theta_D$ is the
Debye temperature):
\begin{equation}
\Delta_{\lambda{\bf k}}=\Delta_{\lambda{\bf k}}^{qh}+
\Delta_{\lambda{\bf k}}^{3}+\Delta_{\lambda{\bf k}}^{4} \label{eq1}
\end{equation}
\begin{equation}
\Delta_{\lambda{\bf k}}^{qh}=-\gamma_{\lambda{\bf k}}\omega_{\lambda{\bf k}}
\frac{\Delta\Omega}{\Omega}, \label{eq2}
\end{equation}
\begin{equation}
\Delta_{\lambda{\bf k}}^{3}=
-\frac{T}{4 M^3 \omega_{\lambda{\bf k}}}
\sum_{\mu\nu{\bf q}} \left\vert V_{\lambda\mu\nu}^{{\bf k},{\bf q},{\bf k}+
{\bf q}} \right \vert^2 \frac{1}{\omega_1^2 \omega_2^2}
\frac{(\omega_1^2-\omega_2^2)^2-\omega_{\lambda{\bf k}}^2
(\omega_1^2+\omega_2^2)}{(\omega_1^2+\omega_2^2-\omega_{\lambda{\bf k}}^2)^2
-4\omega_1^2\omega_2^2}, \label{eq3}
\end{equation}
\begin{equation}
\Delta_{\lambda{\bf k}}^{4}=\frac{T}{4 M^3 \omega_{\lambda{\bf k}}}
\sum_{\mu{\bf q}}  W_{\lambda\lambda\mu\mu}^{{\bf k},{\bf k},{\bf q},
{\bf q}} \frac{1}{\omega_{\mu{\bf q}}^2}, \label{eq4}
\end{equation}
\begin{equation}
\Gamma_{\lambda{\bf k}}=T\frac{\pi}{8 M^3}
\sum_{\mu\nu{\bf q}} \left\vert V_{\lambda\mu\nu}^{{\bf k},{\bf q},{\bf k}+
{\bf q}} \right \vert^2 \frac{1}{\omega_1^2 \omega_2^2}
[\delta(\omega_{\lambda{\bf k}}-\omega_1-\omega_2)+2\delta
(\omega_{\lambda{\bf k}}+\omega_1-\omega_2)] \label{eq5}
\end{equation}
Here $\lambda$, $\mu$, and $\nu$ are the indices of the phonon branches,
{\bf k} and {\bf q} are quasimomenta, $\omega_{\lambda{\bf k}}$,
$\Delta_{\lambda{\bf k}}$, and $\Gamma_{\lambda{\bf k}}$ are, respectively,
the initial phonon frequency and its shift and the phonon damping, $\hat V$
and $\hat W$ are the amplitudes of three- and four-phonon processes (see
Ref. 4), $\gamma_{\lambda{\bf k}}=-\partial\omega_{\lambda{\bf k}}/
\partial\Omega$ is Gruneisen parameter. $\Delta\Omega$ is the change in the
volume $\Omega$ per atom due to thermal expansion, and $1\equiv(\mu, {\bf q})$
and $2\equiv (\nu, {\bf k}+{\bf q})$.

The basic computational results are presented in Figs. 1-5. It is evident from
Figs. 1 and 2 that the high-temperature asymptotic representations (3)-(5)
obtain very early (for $T\gtrsim\Theta_D/3$) Comparing the analogous results
for bcc metals$^4$ it can be concluded that this property is probably quite
general. It follows from Figs. 1-4 that the strongest AEs in Ir occur for
longitudinal phonons near the point X. Specifically, the decrease of the
frequency and the increase of the damping with temperature are greatest for
these phonons The situation differs sharply in this respect from bcc alkali and
alkaline-earth metals, Refs. 4,7 where the maximum damping obtains for "soft"
phonons, for which the opposite
behavior of the frequency with increasing $T$ is characteristic --- hardening
instead of softening. It is evident from Fig. 5 that this difference is due to
the fact that in Ir three-phonon processes dominate over four-phonon processes.

To understand the role of the characteristic features of the interatomic
interactions for the same lattice geometry, the computational results tor
a hypothetical fcc phase of K are presented in Figs. 3 and 4 The explicit
domination of three-phonon over four phonon processes, which causes the
frequency shift to be negative over the entire Brillouin zone is also seen in
this case and is apparently characteristic for all metals with  fcc structure
At the same time, the fact that the neighborhood of the point X is
distinguished probably occurs only for crystals with a "hard" interatomic
interaction potential of the type in Ir.

It is well known Ref. 8 that the "hardness" of the potential in metals
increases
with the effective valence Z (Z= 1 for K and Z=4.5 for Ir). In any case Ir is
a striking example of a crystal where AEs are maximum for high-frequency
vibrations, in contrast to crystals with soft modes where the AEs are maximum
for low-frequency phonons. The latter can be easily explained by the high
powers of the frequency in the denominators of the Eqs. (3)-(5), but the
example of Ir shows that in the general case the question cannot be solved on
the basis of such simple considerations.

According to Fig. 4, an important feature of phonon damping is its nonmonotonic
dependence on the wave vector in the $\Gamma-L$ and $\Gamma-X$ directions.
According to the experimental data presented in Ref. 5, such nonmonotonic
behavior is observed in Al, Cu, Ag, and Au. It can therefore be supposed that
this behavior is typical for all metals with the fcc structure.

In closing we note that, as follows from Figs. 3-5, the AEs have a strong
dependence over the Brillouin zone. For this reason, approximations which
take them into account "on average" (approximations of the type used in the
recently published Ref. 10), are dangerous. Finally, from the fact, found in
the present work, that three-phonon processes predominate over four-phonon
processes it follows that the well-known approaches such as the self-consistent
phonon approach are inapplicable to metals with a close-packed
structure Ref. 11. 
\vskip 5mm

\underline{\bf references} \vskip 3mm

\parindent=0mm
1. V. G. Vaks, \underline{Introduction to the Microscopic Theory of
Ferroelectrics}
(Moscow, Nauka, 1973).

2. V. G. Vaks, S. P. Kravchuk, and A V Trefilov, 
\underline{J. Phys.} 10(F) (1980), 2325.

3. M. I. Katsnel'son and A. V. Trefilov,
\underline{Fiz. Met. Metalloved.} 64 (1987), 629. 

4. V. G. Vaks, S. P. Kravchuk, and A. V. Trefilov,
\underline{J. Phys.}, 10(F) (1980), 2105.

5. M. Zoli, G. Santoro, V. Bortolani et al., 
\underline{Phys. Rev.}, 41(B) (1990), 7507.

6. Y. Y. Ye, Y. Chen, K. M. Ho el al,   
\underline{Phys. Rev. Lett.} 58 (1987), 1769;
\underline{J. Phys.: Condens. Matter} 3 (1991), 9629.

7. V. G. Vaks, G. D. Samolyuk, and A. V. Trefilov,
"On anomalies of anharmonic effects for soft phonons in alkali and bcc
alkaline-earth metals",
\underline{Phys. Lett.}, 127(A) (1988), 37.

8. A. S. Ivanov, M. I. Katsnelson, A. G. Mikhin et al.,
"Phonon spectra, interatomic interaction potentials and simulation
of lattice defects in iridium nad rhodium",
\underline{Philos. Mag.}, 69(B) (1994), 1183.

9. M. I. Katsnelson, I. I. Naumov, A. V Trefilov et al.,
"Peculiarities of phonon spectra and lattice heat capacity in Ir and Rh",
\underline{Philos. Mag.}, 75(B) (1997), 389.

10 R. C. Shukla and E. R. Cowley,         
\underline{Phys. Rev.}, 58(B) (1998), 2596.

11. H. Bottger,
\underline{Principles of the Theory of Lattice Dynamics}
(Weinheim, Physik-Verlag, 1983; Moscow, Mir, 1986).

\vskip 5mm
{\bf Figure captions}
\begin{enumerate}
\item Temperature dependence of the frequency shift in Ir at the symmetric
points of the Brillouin zone; $T_m$
--- melting temperature, $\Theta_D$ --- Debye temperature.
The indices 1 and 2 denote longitudinal and transverse
phonons; $\omega_{pl}$ --- ion plasma frequency
($\omega_{pl}^{Ir}=871$ K, see Ref. 8).
\item Temperature dependence of phonon damping in Ir. All notations are the
same as in Fig 1.
\item Dispersion of the relative frequency shift $\delta= \Delta/\omega$ at
$T= T_m$. Solid line - Ir, dotted line - fcc K. The
numbers 1, 2 indicate longitudinal and transverse branches.
\item Dispersion of the relative phonon damping
$\eta= \Gamma/\omega$ at $T= T_m$. Solid line - Ir, dotted line - fcc K.
The numbers 1, 2 indicate longitudinal and transverse branches.
\item Variation of the anharmonic contributions
$\Delta^{qh}$ (a), $\Delta^3$ (b), and $\Delta^4$ (c) in the
$\Gamma - X - W$ directions in Ir
at $T= T_m$.
\end{enumerate}

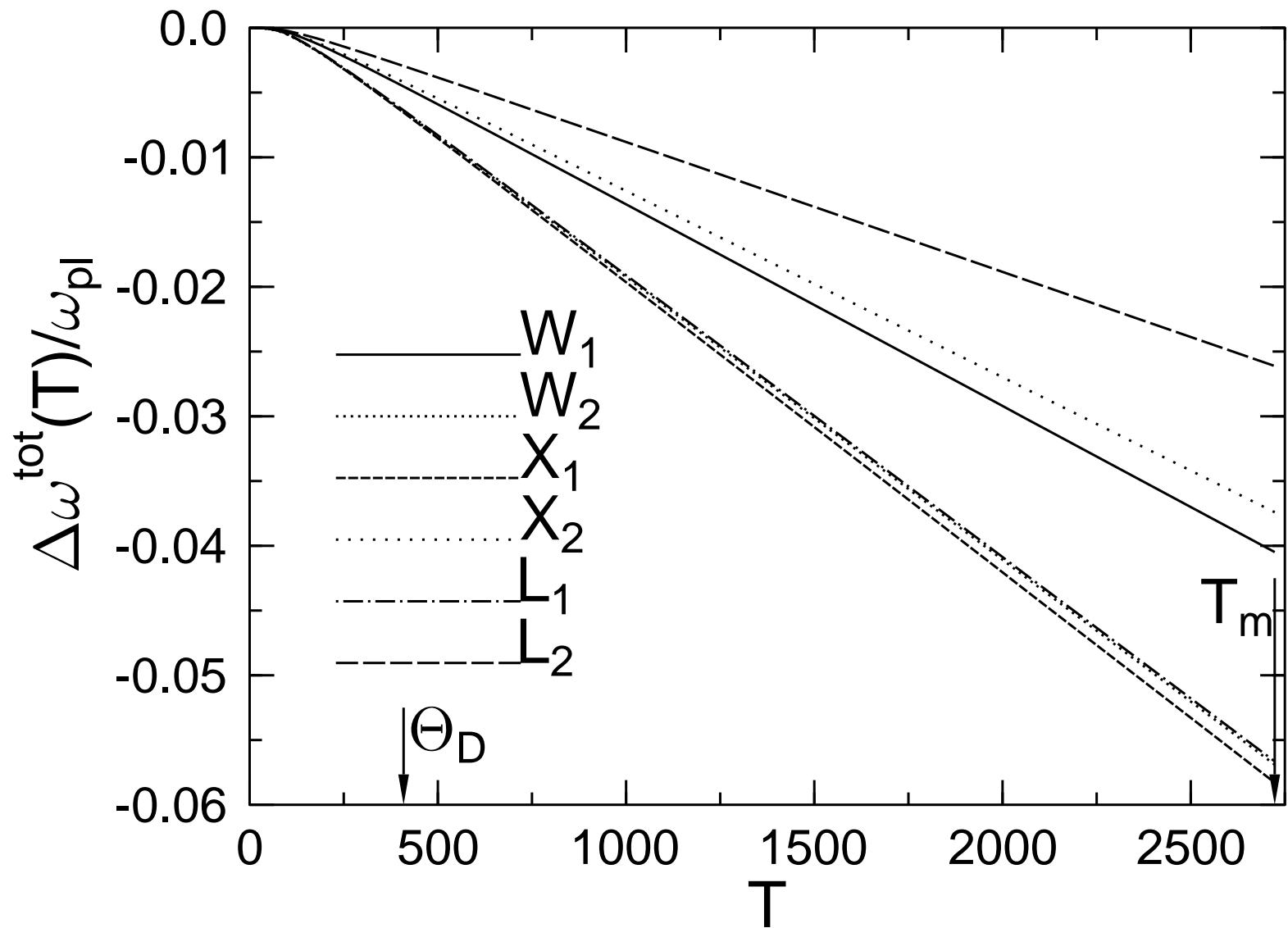
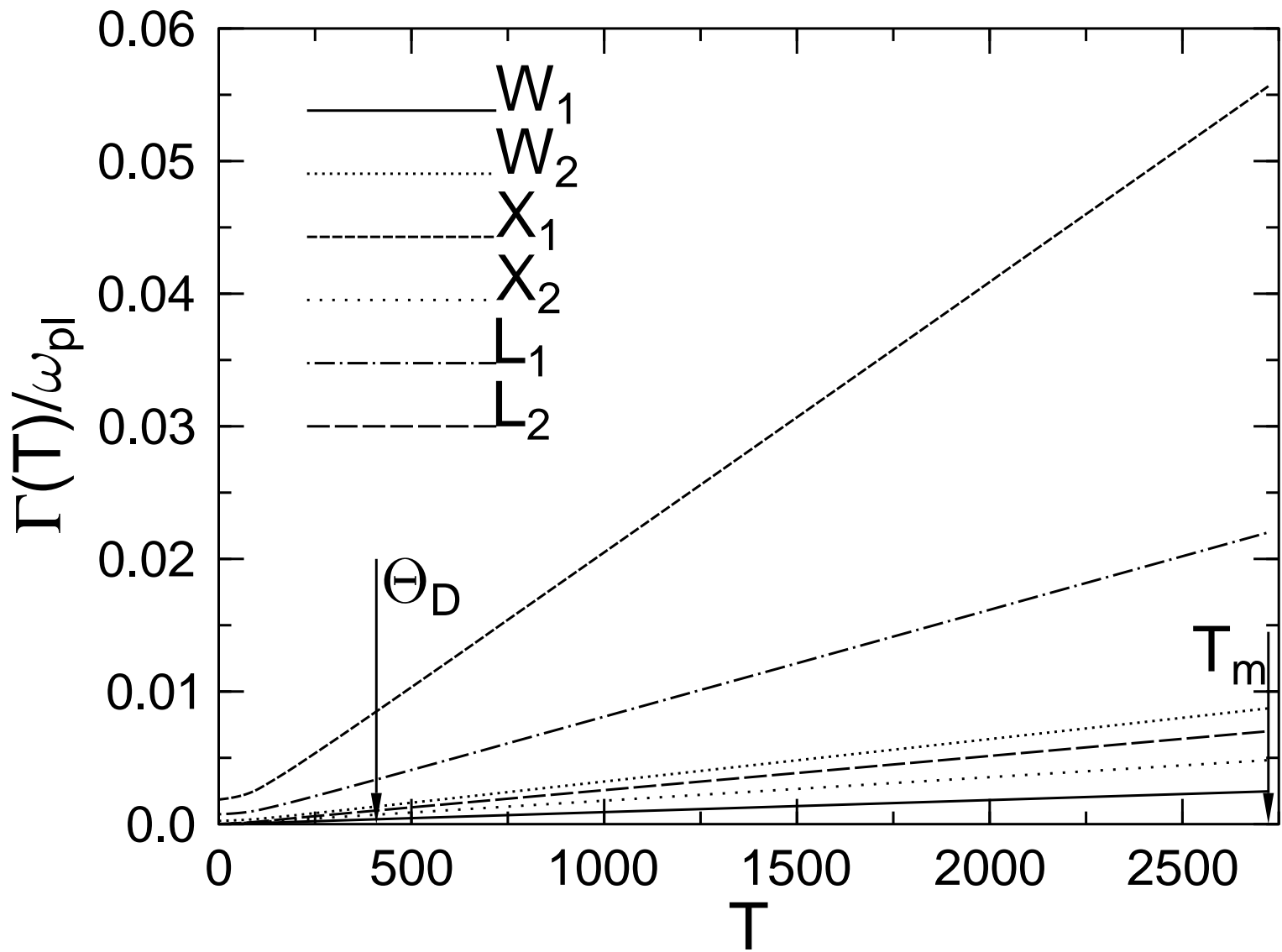
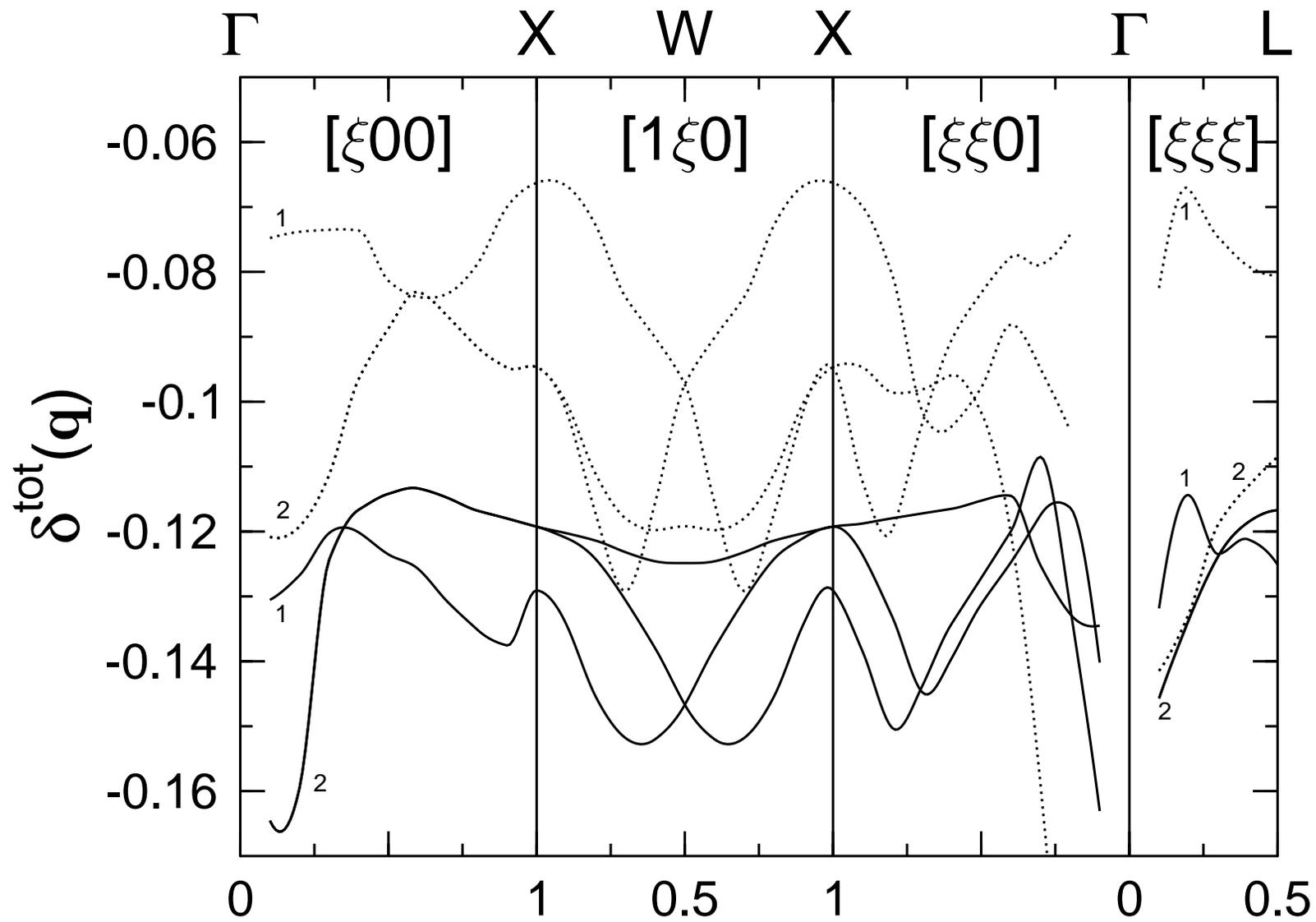
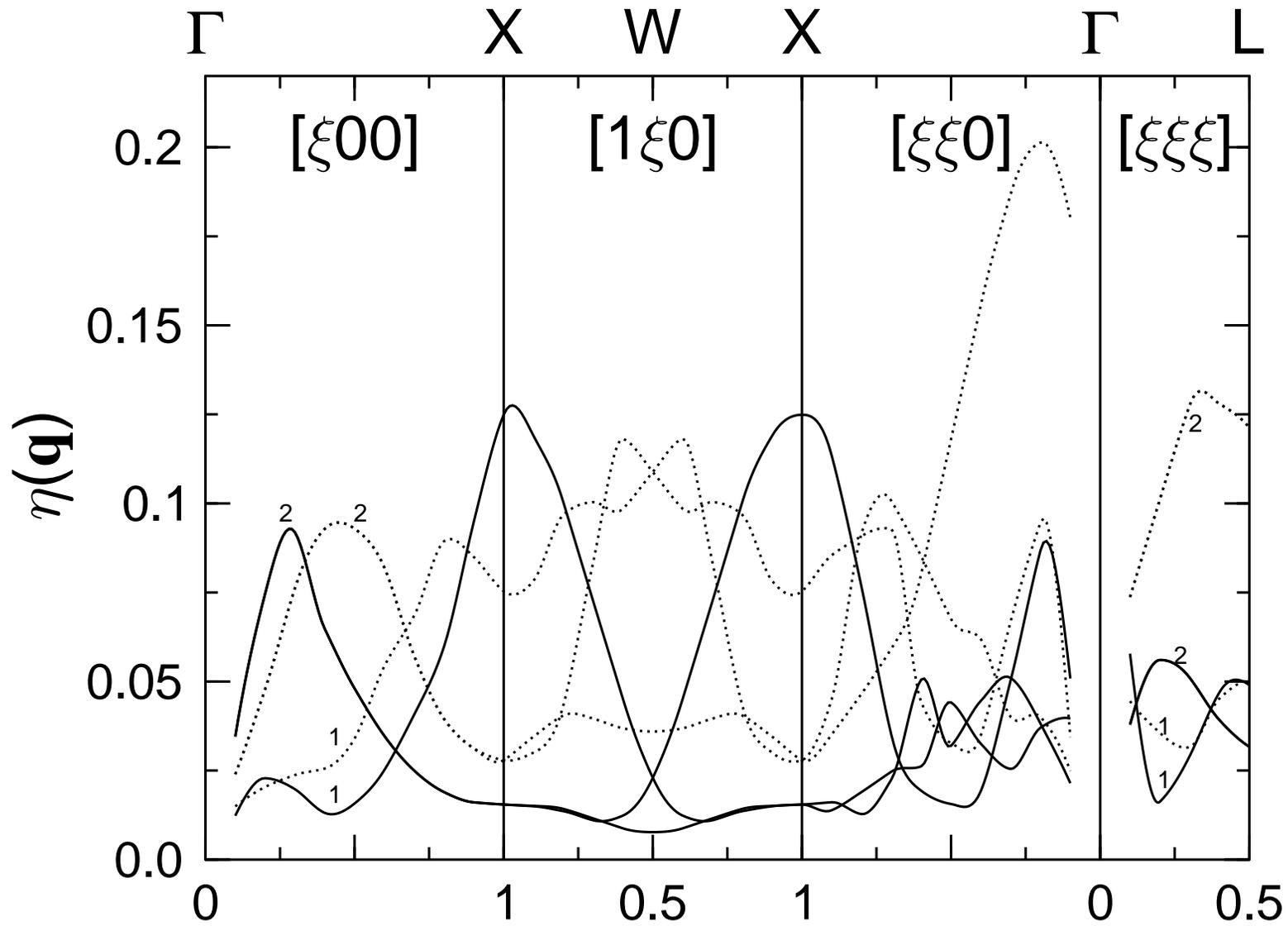
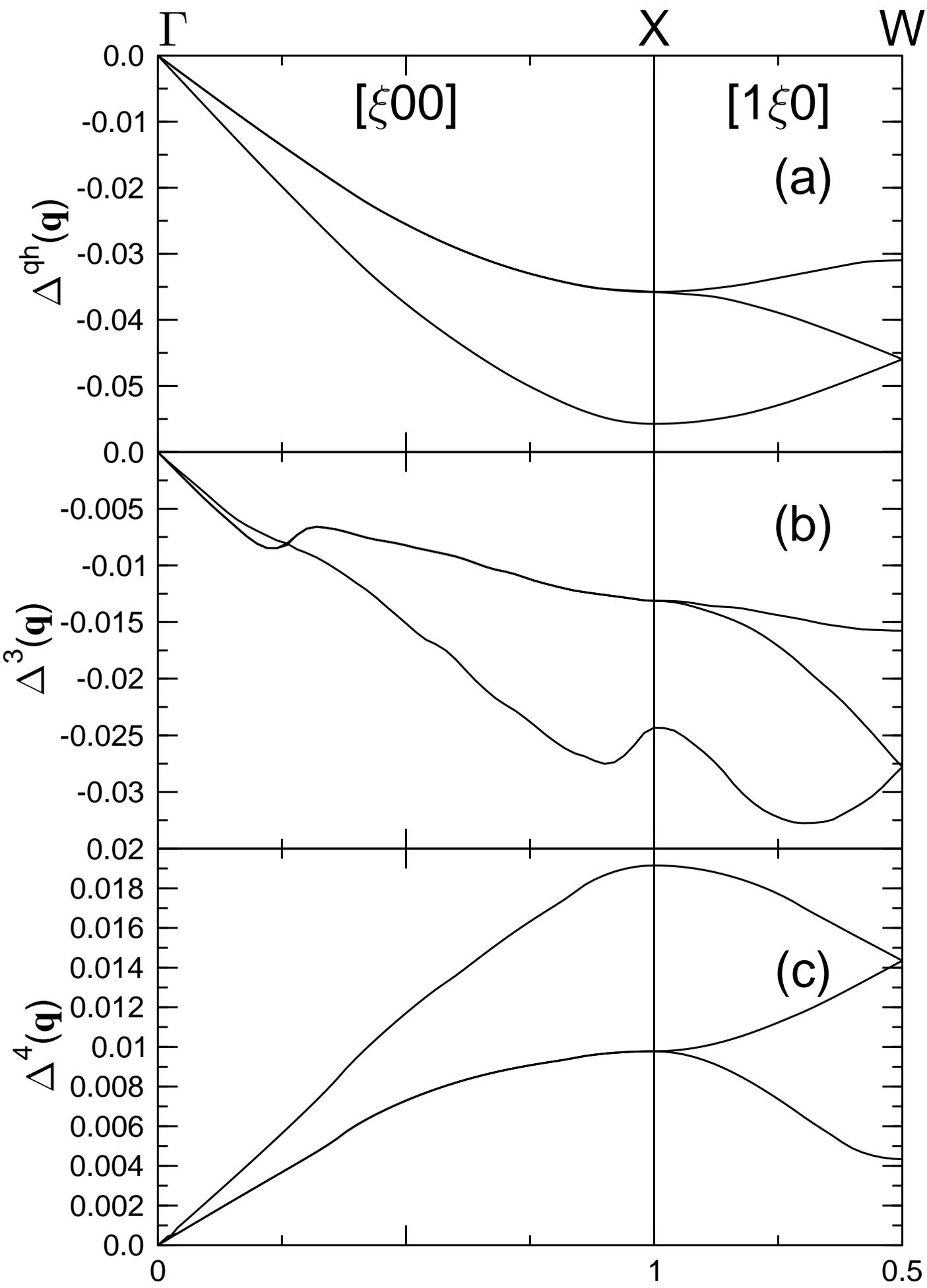
\end{document}